\documentclass{pnastwo}
\usepackage{nidanfloat}

\newcommand{\R}{\mathbb{R}}

\newcommand{\cA}{\mathcal{A}}

\newcommand{\cAamorphous}{\mathcal{A}_{\rm amo}}
\newcommand{\cAliquid}{\mathcal{A}_{\rm liq}}

\newcommand{\bv}[1]{{\boldsymbol #1}}

\begin{document}

\title{Hierarchical structures of amorphous solids characterized by persistent homology}

\author{
Yasuaki Hiraoka\affil{1}{WPI-AIMR, Tohoku University, Japan},
Takenobu Nakamura\affil{1}{},
Akihiko Hirata\affil{1}{},
Emerson G. Escolar\affil{2}{Graduate School of Mathematics, Kyushu University, Japan},
Kaname Matsue\affil{3}{The Institute of Statistical Mathematics, Japan},
\and
Yasumasa Nishiura\affil{1}{}}

\significancetext{Persistent whomology is an emerging mathematical concept for characterizing shapes of data. In particular, it provides a tool called the persistence diagram that extracts multi-scale topological features such as rings and cavities embedded in atomic configurations. This article presents a unified method using persistence diagrams for studying the geometry of atomic configurations in amorphous solids. The method highlights  hierarchical structures that conventional techniques could not have  treated appropriately. 
}

\maketitle

\begin{article}
\begin{abstract} 
This article proposes a topological method that extracts hierarchical structures of various amorphous solids. 
The method is based on the persistence diagram (PD), a mathematical tool for capturing shapes of multi-scale data. 
The input to the PDs is given by an atomic configuration and the output is expressed as two dimensional histograms. 
Then, specific distributions such as curves and islands in the PDs identify meaningful shape characteristics of the atomic configuration. 
Although the method can be applied to a wide variety of disordered systems, it is applied here to silica glass, the Lennard-Jones system, and Cu-Zr metallic glass as standard examples of continuous random network and random packing structures. 
In silica glass, the method classified the atomic rings as short-range and medium-range orders, and unveiled hierarchical ring structures among them. These detailed geometric characterizations clarified a real space origin of the first sharp diffraction peak, and also indicated that PDs contain information on elastic response.
Even in the Lennard-Jones system and Cu-Zr metallic glass, the hierarchical structures in the atomic configurations were derived in a similar way using PDs, although the glass structures and properties substantially differ from silica glass. 
These results suggest that the PDs provide a unified method that extracts greater depth of geometric information in amorphous solids than conventional methods.
\end{abstract}


\abbreviations{PD, persistence diagram; SRO, short-range order; MRO, medium-range order; FSDP, first sharp diffraction peak}

\dropcap{T}he atomic configurations of amorphous solids are difficult to characterize. 
Because they  have no periodicity as found in crystalline solids, only local structures have been analyzed in detail. 
Although short-range order (SRO) defined by the nearest neighbor is thoroughly studied,
it is not sufficient to fully understand the atomic structures of amorphous solids. 
Therefore, medium-range order (MRO) has been discussed to properly characterize amorphous solids \cite{Bernal1964,Polk1971,Finney1970}. 
Many experimental and simulation studies \cite{Elliot1991PRL,ElliotLongman,Susman1991PRB,Greaves2007AdvPhys} have suggested signatures of MRO such as a first sharp diffraction peak (FSDP) 
in the structure factor of the continuous random network structure,
and a split second peak in the radial distribution function
of the random packing structure.
However, in contrast to SRO, the geometric interpretation of MRO and the hierarchical structures among different ranges are not yet clear. 

Among the available methods, the distributions of bond angle and dihedral angle are often used to identify the geometry beyond the scale of SRO.
They cannot, however, provide a complete description of MRO
because they only deal with the atomic configuration up to the third nearest neighbors.
Alternatively, ring statistics are also applied as a conventional combinatorial topological method 
\cite{Polk1971,ISAACS2010,WHZachariasenJACS1932}.
However, this method is applicable only for the continuous random network or crystalline structures, and furthermore, 
cannot describe length scale.
Therefore, methodologies that precisely characterize hierarchical structures beyond SRO and are applicable to a wide variety of amorphous solids are highly desired. 

In recent years, topological data analysis \cite{eh,carlsson} has rapidly grown and has provided
with several tools for studying multi-scale data arising in physical and biological fields
\cite{carlsson,carlssonpnas,Hirata2013,mischaikowprl,NakamuraIOP2015,K}. 
A particularly important tool in the topological data analysis is  persistence diagram (PD), 
a visualization of persistent homology as a two-dimensional histogram (e.g., see Fig.\ \ref{fig:pd_cgl}). 
The input to the PD is given by an atomic configuration with scale parameters, and the output consists of various multi-scale information about topological features such as rings and cavities embedded in the atomic configuration. 
Here, the atomic configurations are generated by molecular dynamics simulations in this article. 
Importantly, in contrast to other topological tools,  PDs not only count topological features, but also provide the scales of these features. 
Hence, PDs can be used to classify topological features by their scales and clarify geometric relationships among them; 
this is presumably the most desired function for deeper analysis of amorphous structures.

This article proposes a method using PDs for various amorphous solids in a unified framework. The method is applied to atomic configurations and enables one to study hierarchical geometry embedded in amorphous structures, which cannot be treated by conventional methods. 
We first applied the method to silica glass as an example of the continuous random network structure,
and obtained the following results:
(1) We found three characteristic curves in the PD of silica glass. These curves classify the SRO rings in the ${\rm SiO}_4$ tetrahedra and the MRO rings constructed by those tetrahedra. Furthermore, a hierarchical  relationship among the SRO and MRO rings was elucidated. 
(2) The PD reproduced the wavelength of the FSDP and clarified a real space origin of the FSDP. 
(3) Each curve in the PD represents a geometric constraint on the ring shapes and, as an example, 
an MRO constraint on rings consisting of three oxygen atoms was explicitly derived as a surface in a parameter space of the triangles. 
Moreover, we verified that these curves are preserved under strain, indicating that the PD properly encodes the material property of elastic response.
Next, as examples of the random packing structure, the Lennard-Jones system and Cu-Zr metallic glass were studied by the PDs, and we clarified the following: 
(4) These amorphous solids were also characterized well by the distributions of curves and islands in the PDs. 
(5) In the Lennard-Jones system, the global connectivity of dense packing regions was revealed by dualizing octahedral arrangements. 
(6) In Cu-Zr alloys, we found that the pair-distribution function defined by the octahedral region in the PD shows the split second peak. Furthermore, a relationship between the hierarchical ring structure and high glass forming ability was discovered in Cu-Zr alloys.

\section{PDs of atomic configuration} 
The input to PDs  is a pair $\cA=(Q,R)$ 
of an atomic configuration $Q=(\bv x_1,\dots,\bv x_N)$ 
and a parameter set $R=(r_1,\dots,r_N)$.
Here, $\bv x_i$ and $r_i$ are 
the position in $\R^3$ and the input radius for the $i$th atom, respectively.
In order to characterize the multi-scale properties in $Q$, 
we introduce a parameter $\alpha$, which controls resolution, and generate a family of atomic balls 
$B_i(\alpha)=\left\{ \bv x \in \R^3\mid ||\bv x-\bv x_i|| \le r_i(\alpha)\right\}$ 
having the radius $r_i(\alpha)=\sqrt{\alpha+r_i^2}$. 
We vary the radii of the atomic balls by $\alpha$ and 
detect rings and cavities at each $\alpha$, 
where $\alpha\geq\alpha_{\rm min}:=-\min\{r_1^2,\dots,r_N^2\}$.

Let $c_k$ be a ring or cavity constructed in the atomic ball model $B(\alpha)=\bigcup_{i=1}^NB_i(\alpha)$ at a parameter $\alpha$. To be more precise, a ring (resp. cavity) here means a generator of the homology $H_1(B(\alpha))$ (resp. $H_2(B(\alpha))$) with a field coefficient \cite{hatcher}. 
Then, we observe that there is a value $\alpha=b_k$ (resp. $\alpha=d_k$) 
at which $c_k$ first appears (resp. disappears) in the atomic ball model. 
The values $b_k$ and $d_k$ are called the birth and death scales of $c_k$, respectively. 
The collection of all the $(b_k,d_k)\in \R^2$ of rings (resp. cavities) 
is the PD denoted by $D_1(\cA)$ (resp. $D_2(\cA)$) for $\cA$ (see Fig. \ref{fig:pointcloud}).
It follows from the structure theorem of persistent homology \cite{carlsson} that the PDs are uniquely defined from the input. From this construction, $(b_k,d_k)$ encodes certain scales of each $c_k$. 
For example, in $D_1(\cA)$, $b_k$ indicates the maximum distance between two adjacent atoms in the ring $c_k$, whereas $d_k$ indicates the size of $c_k$.

In this article, our basic strategy is that we transform a complicated atomic configuration into PDs and try to identify meaningful shape information from specific distributions such as curves or islands in the PDs. 
Namely, we reconstruct characteristic atomic subsets from each distribution. 
To this aim, we compute the optimal cycle for each point $(b_k,d_k)\in D_\ell(\cA)$ on the distribution.
Mathematically speaking, for a given homology generator $c_k$ of $(b_k,d_k)$, the optimal cycle is obtained by solving a minimizing problem in the representatives of $c_k$ under $\ell_1$-norm (see \cite{optimalcycle,op_eh}). 
Our method combining PDs with optimal cycles provides a tool to study inverse problems of PDs, and effectively works in the geometric analysis of glass structures, as we see shortly.

For a mathematically rigorous introduction of these concepts, see Supporting Information or \cite{eh,carlsson}. 
In this article, the PDs are computed by CGAL \cite{cgal} and PHAT \cite{phat}.

\section{PDs for continuous random network structure}
\begin{figure*}[t]
\begin{center}
\includegraphics[height=4.3cm,width=18cm]{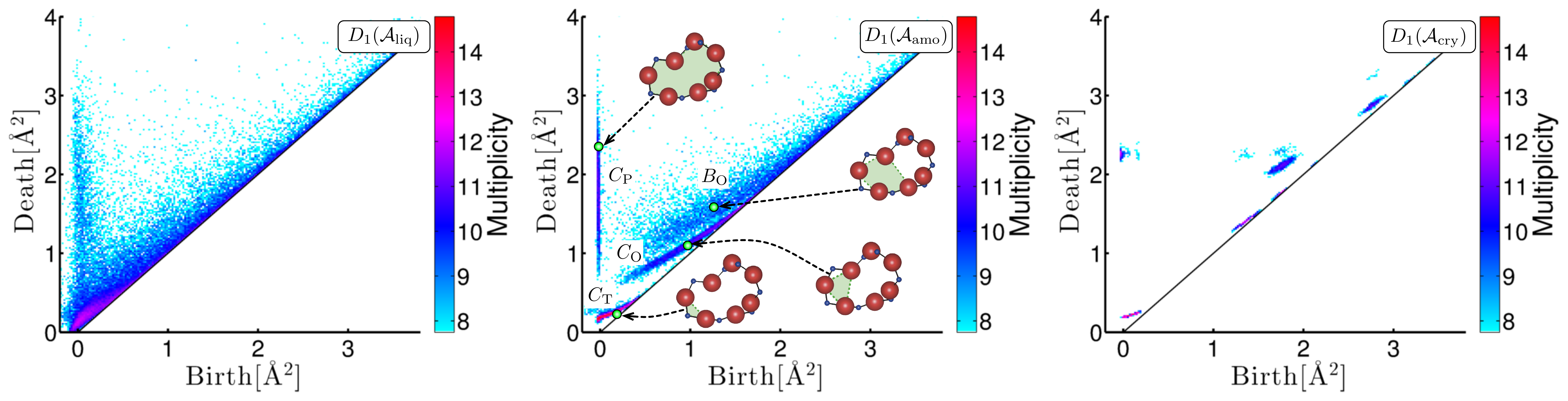}
\caption{PDs of the liquid (left), amorphous (middle), and crystalline (right) states with the multiplicity on the logarithmic scale. In the amorphous state, the three characteristic curves and one band region are labeled $C_{\rm P}, C_{\rm T}, C_{\rm O}$, and $B_{\rm O}$, respectively. 
The insets in $D_1(\cAamorphous)$ show rings in the hierarchical relationship, where
the red and blue spheres represent oxygen and silicon atoms, respectively.
}
\label{fig:pd_cgl}
\end{center}
\end{figure*}

Fig.\ \ref{fig:pd_cgl} shows the PDs 
$D_1(\cA_{\rm liq})$, $D_1(\cA_{\rm amo})$, and 
$D_1(\cA_{\rm cry})$
of a liquid $\cAliquid=(Q_{\rm liq},R_{\rm liq})$,
an amorphous 
$\cA_{\rm amo}=(Q_{\rm amo},R_{\rm amo})$, and  
a crystalline $\cA_{\rm cry}=(Q_{\rm cry},R_{\rm cry})$ state of silica, respectively. 
Here, the horizontal and vertical axes are the birth and death scales, respectively, and the multiplicity of the PDs  is plotted on a logarithmic scale. 
The configurations $Q_{\rm liq}$, $Q_{\rm amo}$ and $Q_{\rm cry}$ are acquired by molecular dynamics simulations using the BKS model \cite{BKS}.
The input radii $R$ are set to be $r_{\rm O}=1.275$ \AA~and $r_{\rm Si}=0.375$ \AA~for each type of the atom (O or Si), which were determined from the first peak positions of the partial radial distribution functions of the amorphous configuration $Q_{\rm amo}$. 
The details of the molecular dynamics simulations and input radii are given in Supporting Information. 

\begin{figure}[h]
\begin{center}
\includegraphics[height=7cm,width=8.3cm]{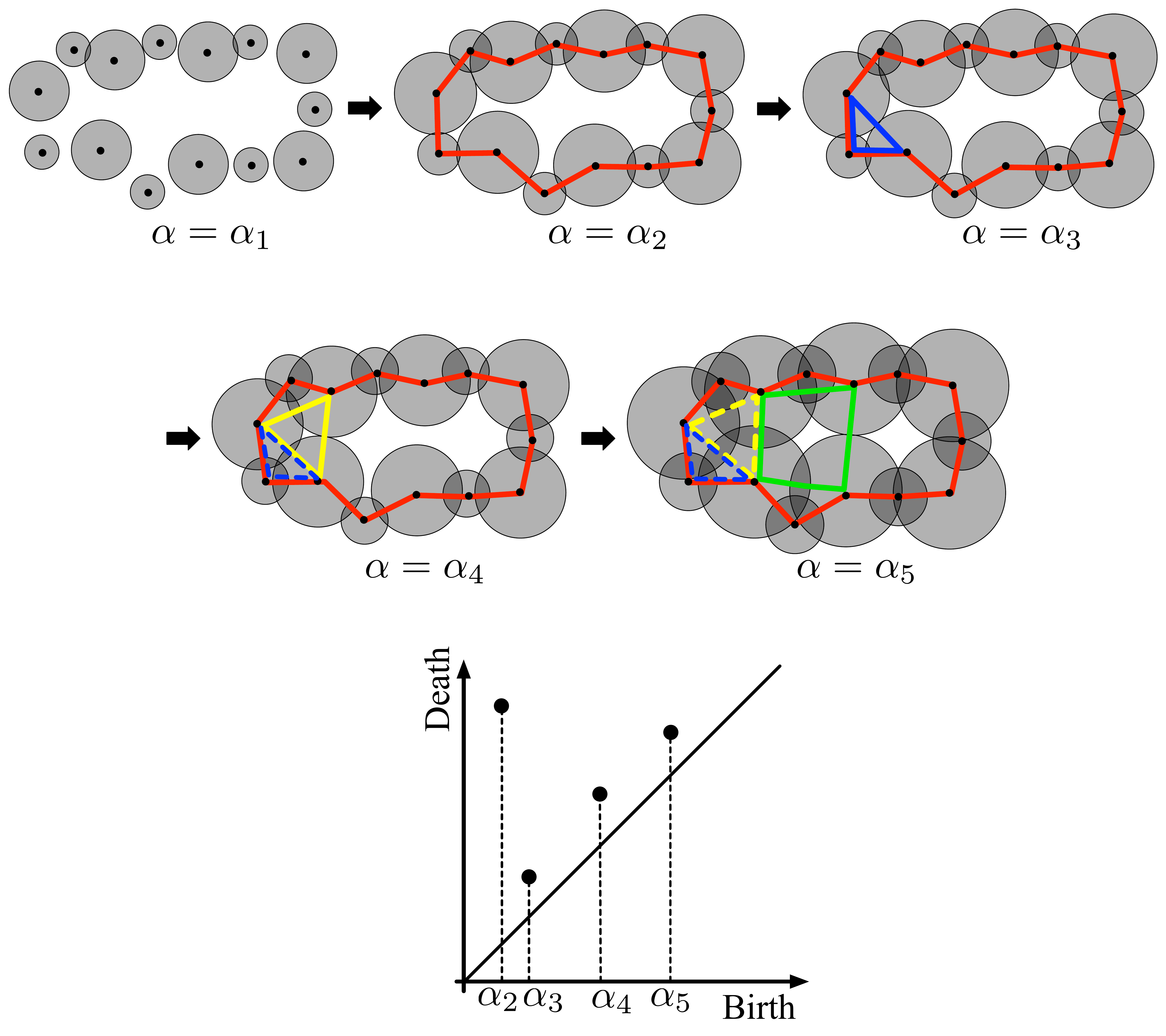}
\end{center}
\caption{Atomic balls (top) and the PD  $D_1(\cA)$ (bottom). New rings appear at $\alpha_i~(i=2,3,4,5)$, and the dashed rings express the disappearance. This is a schematic illustration showing the rings on $C_{\rm P}$ (red), $C_{\rm T}$ (blue), $C_{\rm O}$ (yellow), and $B_{\rm O}$ (green) in silica glass. The large and small balls correspond to oxygen and silicon atoms, respectively. }
\label{fig:pointcloud}
\end{figure}

We discovered that the PDs in Fig.\ \ref{fig:pd_cgl} distinguish these three states.
The liquid, amorphous, and crystalline states are characterized by
planar (2-dim), curvilinear (1-dim), and island (0-dim) regions of the distributions, respectively. 
Here, the 0- and 2-dimensionality of the PDs result
from the periodic and random atomic configurations of the crystalline and liquid states, respectively. 
Furthermore, we emphasize that the presence of the curves in $D_1(\cAamorphous)$ 
clearly distinguishes the amorphous state from the others. This implies that specific 
geometric features of the rings generating the curves in $D_1(\cAamorphous)$ play a significant role for elucidating amorphous states.

As shown in the figure, $D_1(\cAamorphous)$ contains three characteristic curves 
$C_{\rm P}$, $C_{\rm T}$, and $C_{\rm O}$ and one band region $B_{\rm O}$, which are precisely characterized by using the invariance property with respect to the initial radius \cite{NakamuraIOP2015}. 
These particular distributions, especially $C_{\rm O}$, start to become isolated 
near the glass transition temperature $T=T_{\rm g}$ (see Supporting Information).
Through further analysis of the persistent homology using optimal cycles, 
we found the following three geometric characterizations. 
(i) The rings on $C_{\rm P}$ generate secondary rings
on $C_{\rm T}, C_{\rm O}$, and $B_{\rm O}$ (P is named after {\em primary}). 
That is, by increasing the parameter $\alpha$, 
each ring on $C_{\rm P}$ becomes thicker and starts to create new rings by pinching itself, 
and these newly generated rings appear on $C_{\rm T}, C_{\rm O}$, and $B_{\rm O}$ (a schematic illustration of the pinching process is described in Fig. \ref{fig:pointcloud}). 
This indicates 
a hierarchical structure 
from $C_{\rm P}$ to $C_{\rm T}, C_{\rm O}$, and $B_{\rm O}$ in the continuous random network.
An example of the rings in this hierarchical relationship is depicted in the inset of $D_1(\cAamorphous)$.
(ii) The rings on $C_{\rm T}$ are constructed by tetrahedra consisting of four oxygen atoms at the vertices with one silicon atom at the center.
(iii) The rings on $C_{\rm O}$ and $B_{\rm O}$ are constructed only by the oxygen atoms (three and more, respectively). 
Recalling that the death scale indicates the size of rings, the rings on $C_{\rm T}$ are 
classified as SRO, while those on $C_{\rm P}$, $C_{\rm O}$, and $B_{\rm O}$ are classified as MRO.

\section{Decomposition of FSDP}
The FSDP observed in the structure factor $S(q)$ ($q\sim 1.5$-$2$ ${\rm \AA}^{-1}$) has been supposed to be a signature of MRO, but its real space origin is still controversial \cite{Elliot1991PRL}. 
Here, we found that the distributions $C_{\rm P}$, $C_{\rm O}$, and $B_{\rm O}$ of the MRO rings 
reproduce the $q$ values of the FSDP fairly well. Moreover, we classified the MRO rings as a real space origin of the FSDP. 

We first note that the death scales of the rings on $C_{\rm P}$, $C_{\rm O}$, and $B_{\rm O}$ are determined only by the oxygen atoms, and this is directly verified by the PD computation. 
In addition, recall that $\alpha$ is the parameter controlling the radius $r_i(\alpha)=\sqrt{\alpha+r_i^2}$ of  the $i$-th atomic ball $B_i(\alpha)$, 
and the death scale $\alpha=d_k$ indicates the size of the individual ring $c_k$. 
More precisely, the ring $c_k$ disappears at $\alpha=d_k$ by being covered up in the atomic ball model  $\bigcup_{i=1}^NB_i(\alpha)$.  Hence $\ell(d_k)=2\sqrt{d_k+r_{\rm O}^2}$ measures the size of $c_k$ on $C_{\rm P}$, $C_{\rm O}$ 
and $B_{\rm O}$, where $r_{\rm O}$ is the input radius of the oxygen atom.

From the aforementioned argument, we define a distribution
\[
M_{A_i}(q)=\frac{1}{|A_i|}
\sum_{(b_k,d_k)\in A_i}
\delta\left(q-\frac{2\pi}{\ell(d_k)}\right),
\]
where $A_1=C_{\rm P}\cup C_{\rm O}\cup B_{\rm O}$,
$A_2=C_{\rm P}$, $A_3=C_{\rm O}$, and $A_4=B_{\rm O}$, and $|A_i|$ is the number of the elements in $A_i$. 
Here, $\delta$ is the Dirac delta function which is used to count the contribution of each MRO ring in $q$-space.

\begin{figure}[h]
\begin{center}
\includegraphics[height=4.8cm,width=7cm]{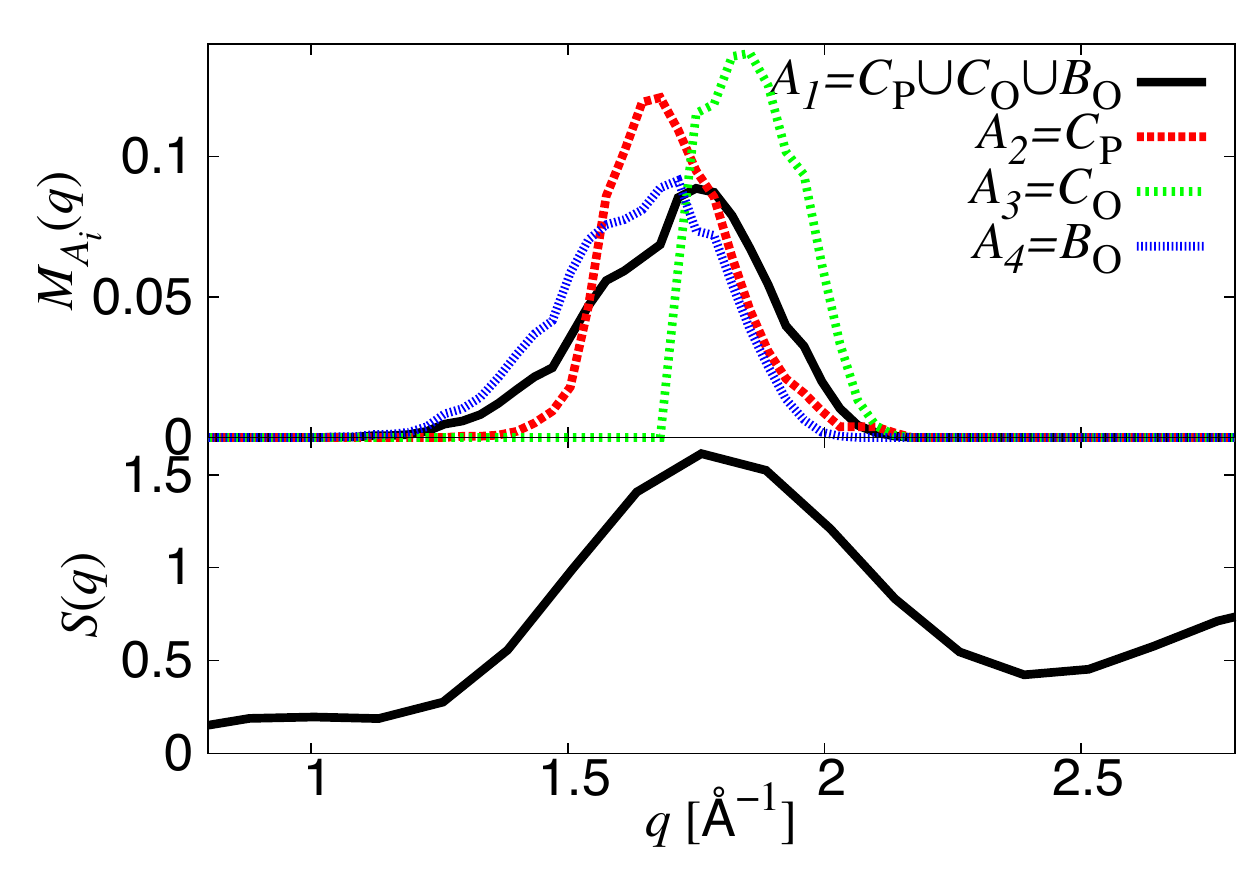}
\end{center}
\caption{The distributions $M_{A_i}(q)$ ($i=1,2,3,4$) and the structure factor $S(q)$ around the FSDP.}
\label{fig:sqcp}
\end{figure}

Fig.\ \ref{fig:sqcp} shows the plots of $M_{A_i}(q)$ and the structure factor $S(q)$ around the FSDP. 
We found good agreement between the $q$ values of the FSDP and the peak of $M_{A_1}(q)$. 
This implies that the MRO rings composed of $C_{\rm O}$, $C_{\rm P}$, and $B_{\rm O}$ are the real space origin of the FSDP. 
We also note that the distributions $M_{A_3}(q), M_{A_2}(q), M_{A_4}(q)$ are located on the large, medium, and small $q$  values and, hence, the rings in $C_{\rm O}$, $C_{\rm P}$, and $B_{\rm O}$ provide a decomposition of the FSDP into those $q$ values, respectively.
It should  be emphasized that the $M_{A_i}(q)$ are derived from the configuration of the oxygen atoms only. This shows that  the configuration of oxygen atoms plays a significant role in the FSDP. 
Furthermore, the invariant property \cite{NakamuraIOP2015} of $\ell(d_k)$ for the rings on $C_{\rm P}$, $C_{\rm O}$ and $B_{\rm O}$ induces that of $M_{A_i}(q)$ under the choice of the input radius $r_{\rm O}$.
This means that our analysis using PDs does not contain any artificial ambiguity of the input radii.

\section{Curves and shape constraints}
The presence of curves $C_{\rm P}$, $C_{\rm T}$, and $C_{\rm O}$ in $D_1(\cAamorphous)$ clearly distinguishes the amorphous state from the crystalline and liquid states. 
We emphasize here that this characteristic property shows 
the constraints on the shapes of the rings induced by the normal directions of these curves.

\begin{figure}[h]
\begin{center}
\includegraphics[height=3.7cm,width=7.5cm]{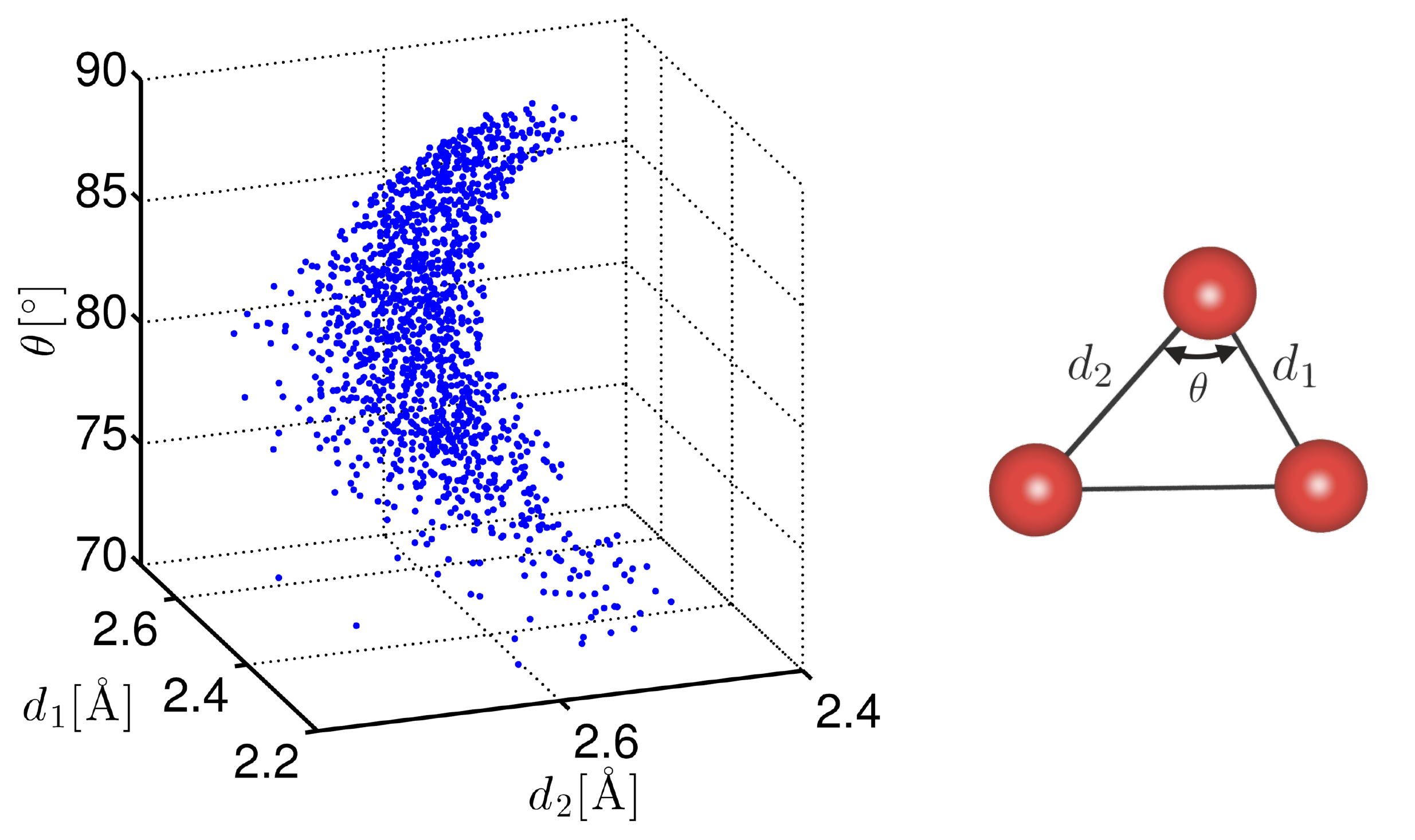}
\end{center}
\caption{Plot of $(d_1,d_2,\theta)$ for the O-O-O triangles on $C_{\rm O}$. }
\label{fig:codim_surface}
\end{figure}

For example, the shape of a ring on $C_{\rm O}$, 
which consists of three oxygen atoms (see the right panel in Fig.\ \ref{fig:codim_surface}), is determined 
by specifying the first and second minimum edge lengths  
$d_1$ and $d_2$ ($d_1<d_2$) and the angle $\theta$ between them, 
and hence is realized in a three dimensional parameter space.
Then, the constraint for $C_{\rm O}$ to be the curve requires that these three variables $(d_1, d_2, \theta)$ satisfy a certain relation $f(d_1,d_2,\theta)=0$,
and hence provides a restriction on the shape of the O-O-O triangles. 
Fig.\ \ref{fig:codim_surface} shows a plot of $(d_1,d_2,\theta)$ for the O-O-O triangles on $C_{\rm O}$ in $D_1(\cAamorphous)$, and we found a surface corresponding to this constraint $f(d_1,d_2,\theta)=0$ in the parameter space.
This demonstrates one of the medium-range geometric structures in the amorphous state. 
It is worth noting that this new characterization of MRO of O-O-O triangles in three dimensional parameter space cannot be derived by separately analyzing the conventional distributions of lengths or angles, since each of them can only deal with the single parameter.

\section{Response under strain}
The presence of the curves also indicates variations 
in the shapes of the rings. 
That is, by following each curve along its tangential direction and
studying its rings, we can observe the deformation of the rings.
It is reasonable to suppose that these variations are due to thermal fluctuations, and hence the deformations of the ring configurations along the curves are probably softer than those in the normal directions. Consequently, the response under strain is expected to follow the same shape constraints.

\begin{figure}[h]
\begin{center}
\includegraphics[height=7.0cm,width=9cm]{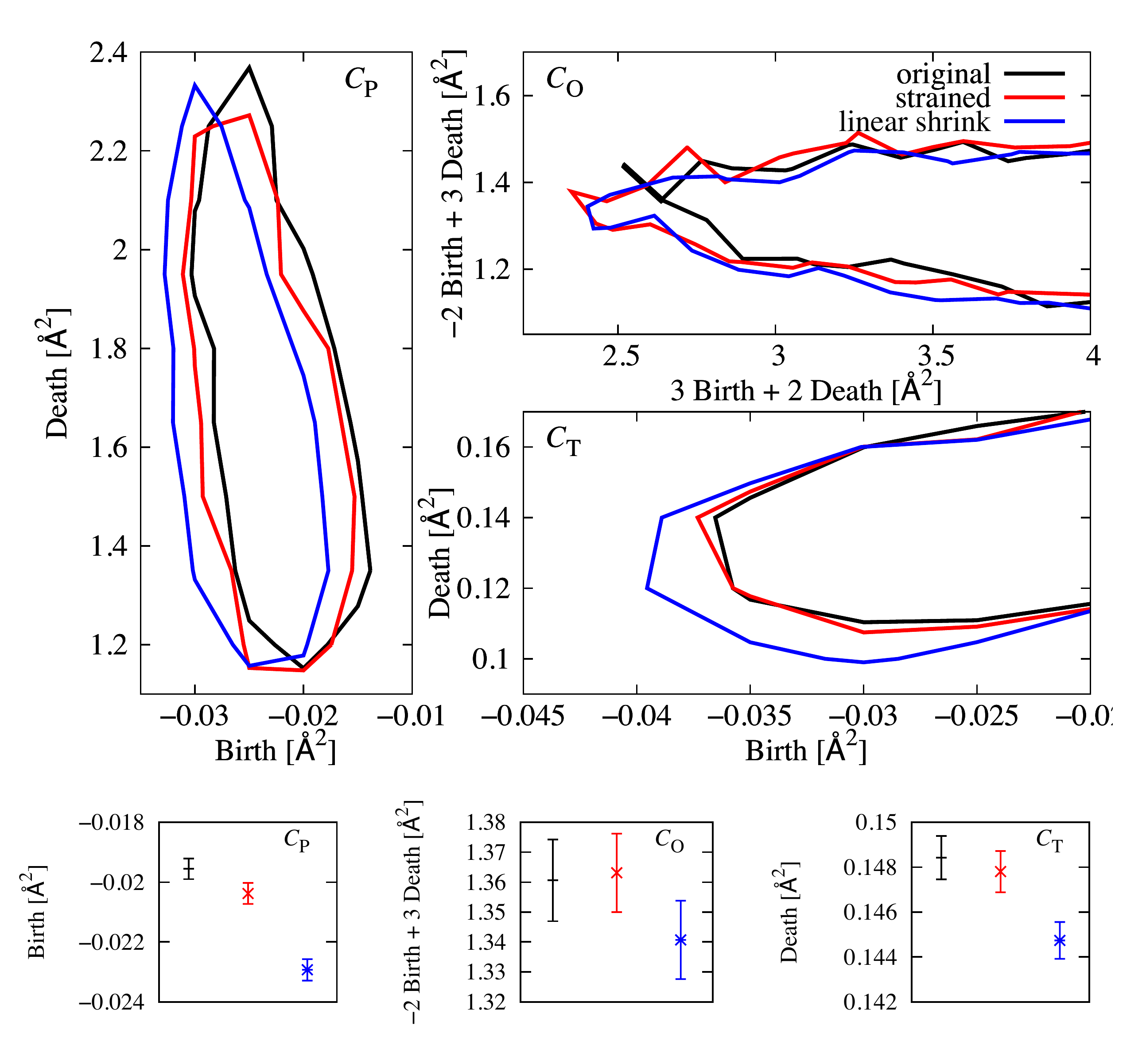}
\end{center}
\caption{Top: Contour plots of $C_{\rm P}$, $C_{\rm O}$, and $C_{\rm T}$ for the original configuration $Q_{\rm amo}$ of the amorphous state (black), the $1\%$ strained state $Q_{\rm amo}^{\rm strain}$ (red), and the artificial $1\%$ linear shrink of the original coordinates $Q_{\rm amo}^{\rm linear}$ (blue), respectively. Bottom: Projections of the histograms on the normal directions.  
}
\label{fig:pressure}
\end{figure}

To verify this mechanical response of PDs, we performed simulations of isotropic compression for our amorphous state, and computed the PD of the strained state. 
The strain is set to be $1\%$ of the original volume,
which is sufficiently small to satisfy a linear response relation with the stress.
The three panels on the top in Fig.\ \ref{fig:pressure} show the contours of the histograms restricted on $C_{\rm P}$, $C_{\rm O}$, and $C_{\rm T}$
for the original configuration $Q_{\rm amo}$ of the amorphous state (black), 
for the strained state $Q_{\rm amo}^{\rm strain}$ (red), and for the artificial $1\%$ linear shrink $Q_{\rm amo}^{\rm linear}$ of the original coordinates (blue), respectively. The contours of $C_{\rm O}$ are depicted on the coordinates along the tangential direction. 
The bottom three panels show the projections of the histograms on their normal axes.

The figures show that the contours of the strained state shift along the original curves $C_{\rm P}$, $C_{\rm O}$, and $C_{\rm T}$ in $D_1(\cAamorphous)$; i.e., downward in $C_{\rm P}$, leftward in $C_{\rm O}$, and almost fixed in $C_{\rm T}$, respectively. 
This is in contrast to a linear shrink, in which the contours simply move in the direction of decreasing both birth and death scales because of the uniform shrink of the system size. 
This strongly suggests that the configuration $Q_{\rm amo}^{\rm strain}$ reflects the shape constraints 
and supports the expected mechanical response of PDs. 
We also note from the figures of $Q_{\rm amo}^{\rm strain}$ that the contour of $C_{\rm T}$ is almost fixed compared with those of $C_{\rm P}$ and $C_{\rm O}$. Since $C_{\rm P}$ and $C_{\rm O}$ ($C_{\rm T}$) are classified as MRO (SRO), this implies that MRO is mechanically softer than SRO. 
Here, we remark that we can observe a similar behavior of the PDs for a shear deformation.

We have revealed that PDs encode information about elastic response, similar to how the radial distribution function encodes volume compressibility \cite{Hansen.SimpleLiquid}. 
It should also be emphasized that the curves or islands appear in the PDs of only the solids.  
This evidence suggests that these isolated distributions are related to the rigidity of the materials. This hypothesis follows from the fact that isolations represent geometric constrains reflecting mechanical responses. Future numerical and theoretical studies to unravel this relationship would be of great value.

\begin{figure}[h]
\begin{center}
\includegraphics[height=6cm,width=8.5cm]{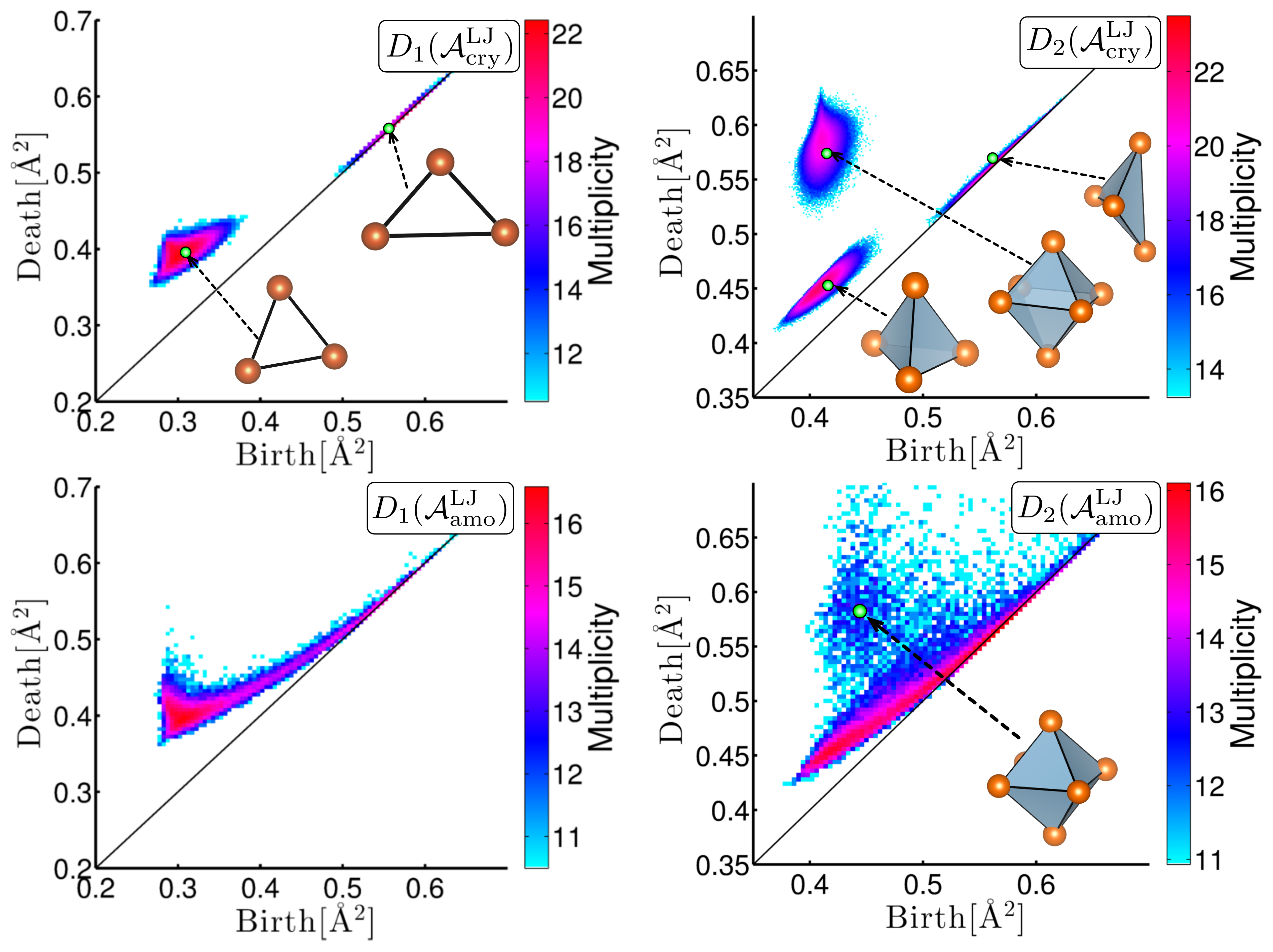}
\end{center}
\caption{PDs for the Lennard-Jones system with the multiplicity on the logarithmic scale. 
The left and right panels correspond to $D_1$ and $D_2$ respectively.
Top panels correspond to FCC crystal at $T=0.1$
and bottom panels correspond to amorphous state at $T=10^{-3}$.
 }
\label{fig:LJ_pd}
\end{figure}

\section{PDs for random packing structures}
We next study the geometry of amorphous states close to random packing structures.
In this case, we found that both $D_1$ and $D_2$ capture characteristic features of the amorphous structures.
Fig.\ \ref{fig:LJ_pd} shows the PDs $D_i(\cA^{\rm LJ}_{\rm cry})$ and
$D_i(\cA^{\rm LJ}_{\rm amo})$ for $i=1,2$ of the mono-disperse Lennard-Jones system 
in the crystalline and the amorphous states, respectively.
The input radii $r=r_1=\dots=r_N$ in $R$ are set to be zero, since changing $r$ only causes translations of the PDs for the single component system. 
The details of the simulation are explained in Supporting Information.

Similar to the case of silica, the crystalline structure is characterized by the island distributions in the PDs (see top panels in Fig.~\ref{fig:LJ_pd}). 
They correspond to the regular triangles in $D_1(\cA^{\rm LJ}_{\rm cry})$ and the regular octahedra, the regular tetrahedra and the quartoctahedra \cite{Gervasi2005} in $D_2(\cA^{\rm LJ}_{\rm cry})$ of the FCC configuration. 
For the amorphous structure, the curves in $D_1(\cA^{\rm LJ}_{\rm amo})$ and $D_2(\cA^{\rm LJ}_{\rm amo})$ represent variations of triangles and tetrahedra, respectively. 
We also note that the isolation of the octahedral distribution is preserved well even in $D_2(\cA^{\rm LJ}_{\rm amo})$. 
Its peak is separated from the curve of the tetrahedra (see bottom right panel in Fig.~\ref{fig:LJ_pd}), 
demonstrating the quantitative classification into two typical local structures of different-sized cavities.

\begin{figure}[h]
\begin{center}
\includegraphics[height=3cm,width=7.7cm]{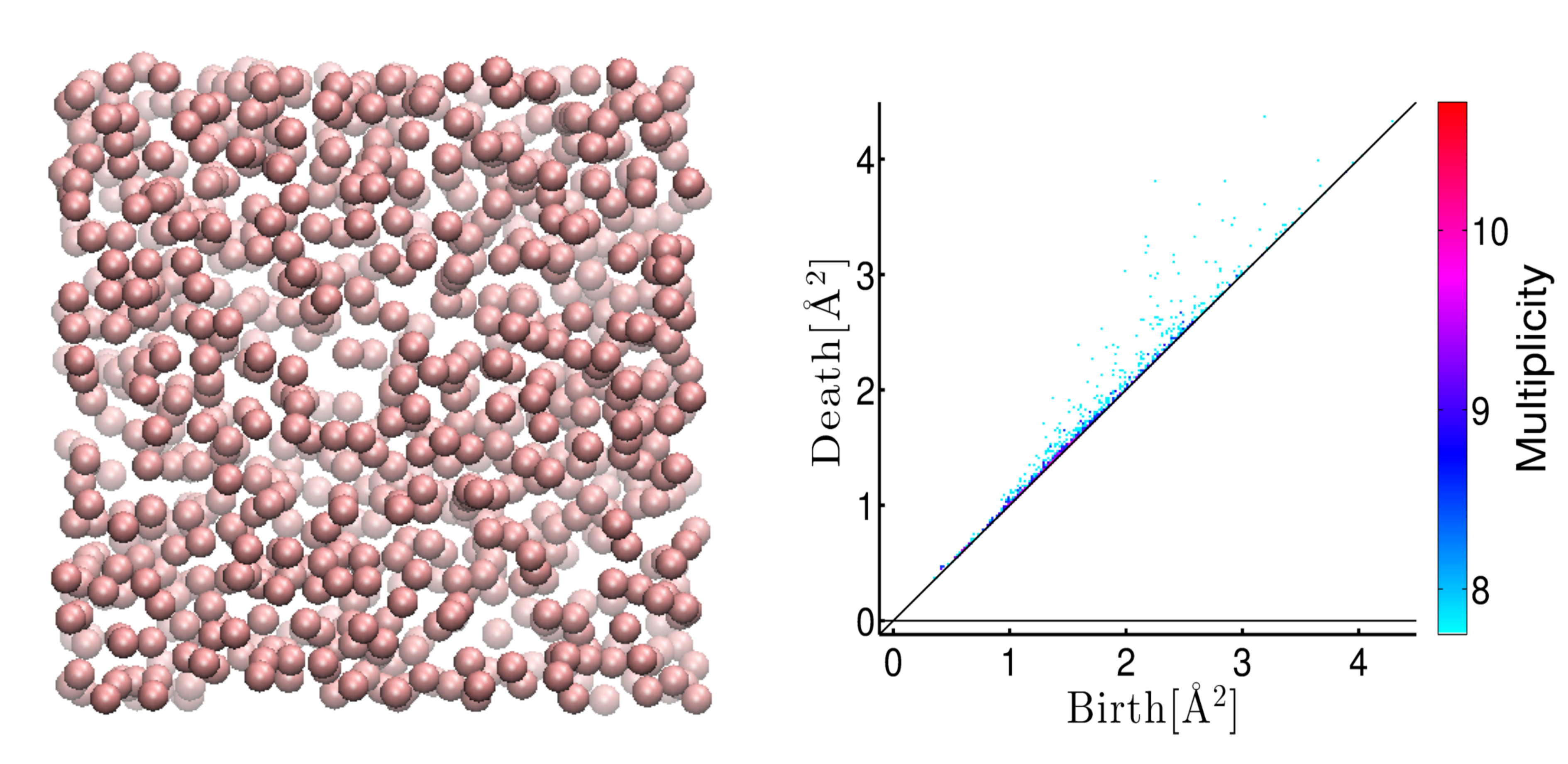}
\end{center}
\caption{Left: Red balls express the octahedra in the amorphous structure of the Lennard-Jones system, and the empty part corresponds to the dense packing regions. Right: PD $D_2(\cA^{\rm LJ}_{\rm oct})$ of the left figure with the multiplicity on the logarithmic scale. 
}
\label{fig:pd2_network}
\end{figure}

In random packings, it is known that the atomic configuration can be divided into dense packing regions built from tetrahedra and the complement that patches those regions together \cite{yamamoto}. 
In particular, the network structure of the dense packing regions characterizes the global connectivity beyond MRO, which has not yet been investigated in detail. 
Note that $D_2(\cA^{\rm LJ}_{\rm amo})$ clearly separates the dense packing regions as  the deformation curve of the tetrahedra and identifies the octahedral island as the main component of the complement structure. From the Alexander duality in $\R^3$ (e.g., \cite{hatcher}), the connectivity of the dense packing regions can be studied by the cavities of the complement.  
Namely, we first extract all octahedra from the octahedral island in $D_2(\cA^{\rm LJ}_{\rm amo})$ and construct a new set of points $\cA^{\rm LJ}_{\rm oct}=(Q^{\rm LJ}_{\rm oct},R^{\rm LJ}_{\rm oct})$ by putting a point at the center of each octahedron (left of Fig. \ref{fig:pd2_network}). Here, we set $R^{\rm LJ}_{\rm oct}$ to be zero, since the standard deviation of the octahedral sizes is very small. 

The right of Fig. \ref{fig:pd2_network} shows $D_2(\cA^{\rm LJ}_{\rm oct})$, and we clearly observe that almost all cavities are located close to the diagonal. This means that the octahedral distribution rarely generates persistent cavities, and hence by the duality, we can conclude that the dense packing regions mostly construct a giant connected network.  
It should be remarked that the dual treatment is much easier and computationally more efficient than directly studying the intricate huge network structures. 
We also emphasize that the analysis here of treating the octahedra as a new input is an iterative usage of the PD method, and can be an effective approach for studying multi-scale geometry.

\begin{figure}[h]
\begin{center}
\includegraphics[height=9.5cm,width=8.5cm]{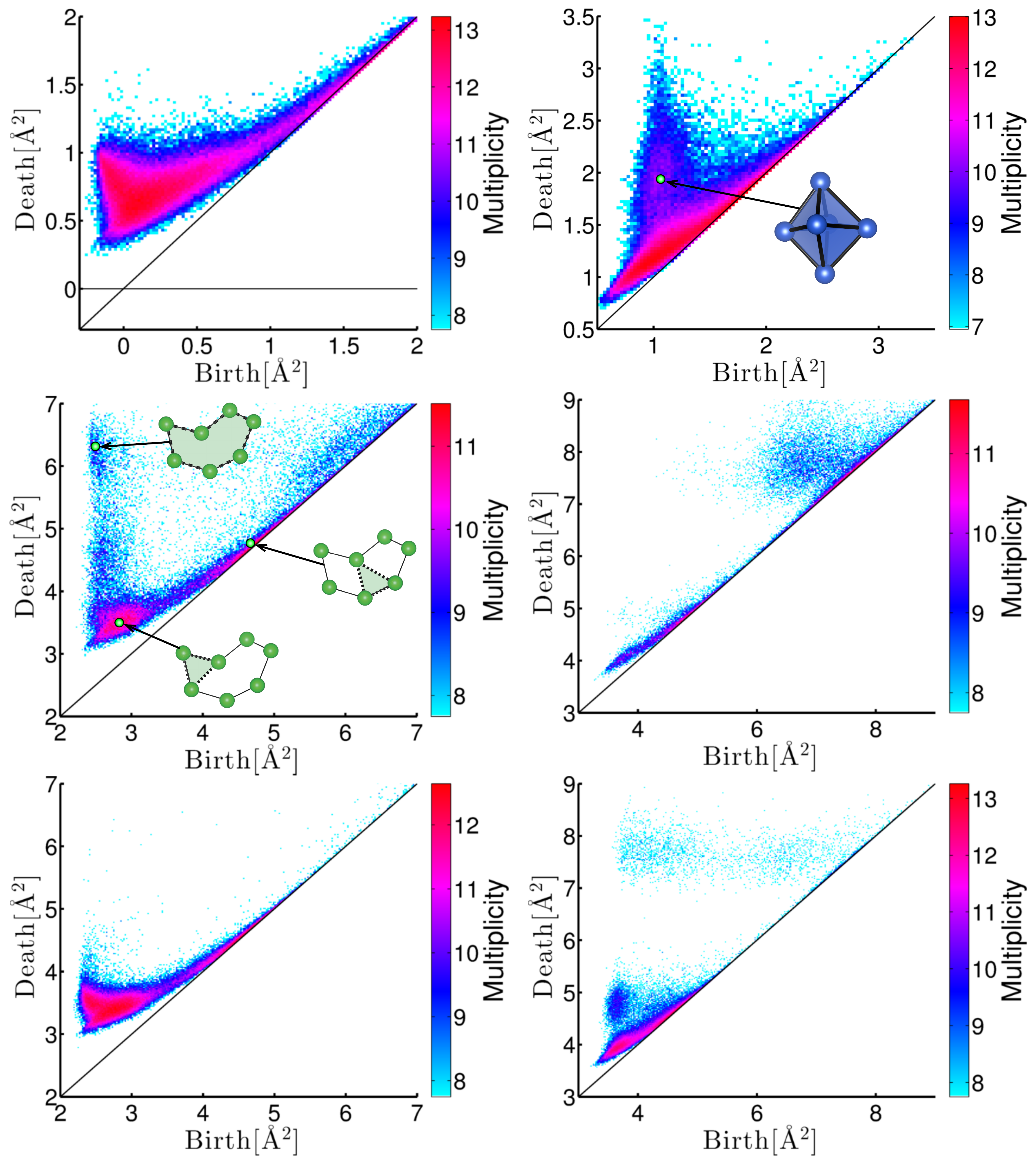}
\end{center}
\caption{
PDs for CuZr alloys with the multiplicity on the logarithmic scale. 
The left and right panels correspond to $D_1$ and $D_2$ respectively.
In the top panels,
$D_i(\cA^{\rm Cu_{50}Zr_{50}}_{\rm amo})$ are described.
The middle and bottom panels 
show the PDs of the atomic configurations of only Zr atoms in the alloys.
The middle panels correspond to ${\rm Cu_{50}Zr_{50}}$ alloy,
and the bottom panels correspond to ${\rm Cu_{15}Zr_{85}}$ alloy.
The blue and green spheres represent copper and zirconium atoms, respectively.}
\label{fig:CuZr_pd}
\end{figure}

As an example of multi-component systems, we also studied
metallic glasses composed of Cu and Zr  \cite{Pan2011}, in particular, focusing on ${\rm Cu_{50}Zr_{50}}$ and ${\rm Cu_{15}Zr_{85}}$ that display the different glass forming ability. 
The PDs for these alloys are shown in Fig.~\ref{fig:CuZr_pd}. 
The input radii for Cu and Zr are set to be 1.30 \AA~and 1.55 \AA,~respectively,
for the multi-component PDs (top panels in Fig.~\ref{fig:CuZr_pd}). 
These values are obtained by the same procedure as for the silica.
Even in the multi-component system, the PDs basically show 
similar behaviors to those of the Lennard-Jones system.
Specifically, the island distribution corresponding to octahedra 
appears in $D_2(\cA^{\rm Cu_{50}Zr_{50}}_{\rm amo})$ and the characteristic curves are also observed in $D_1(\cA^{\rm Cu_{50}Zr_{50}}_{\rm amo})$ and $D_2(\cA^{\rm Cu_{50}Zr_{50}}_{\rm amo})$.

\begin{figure}[h]
\begin{center}
\includegraphics[height=4.2cm,width=7.5cm]{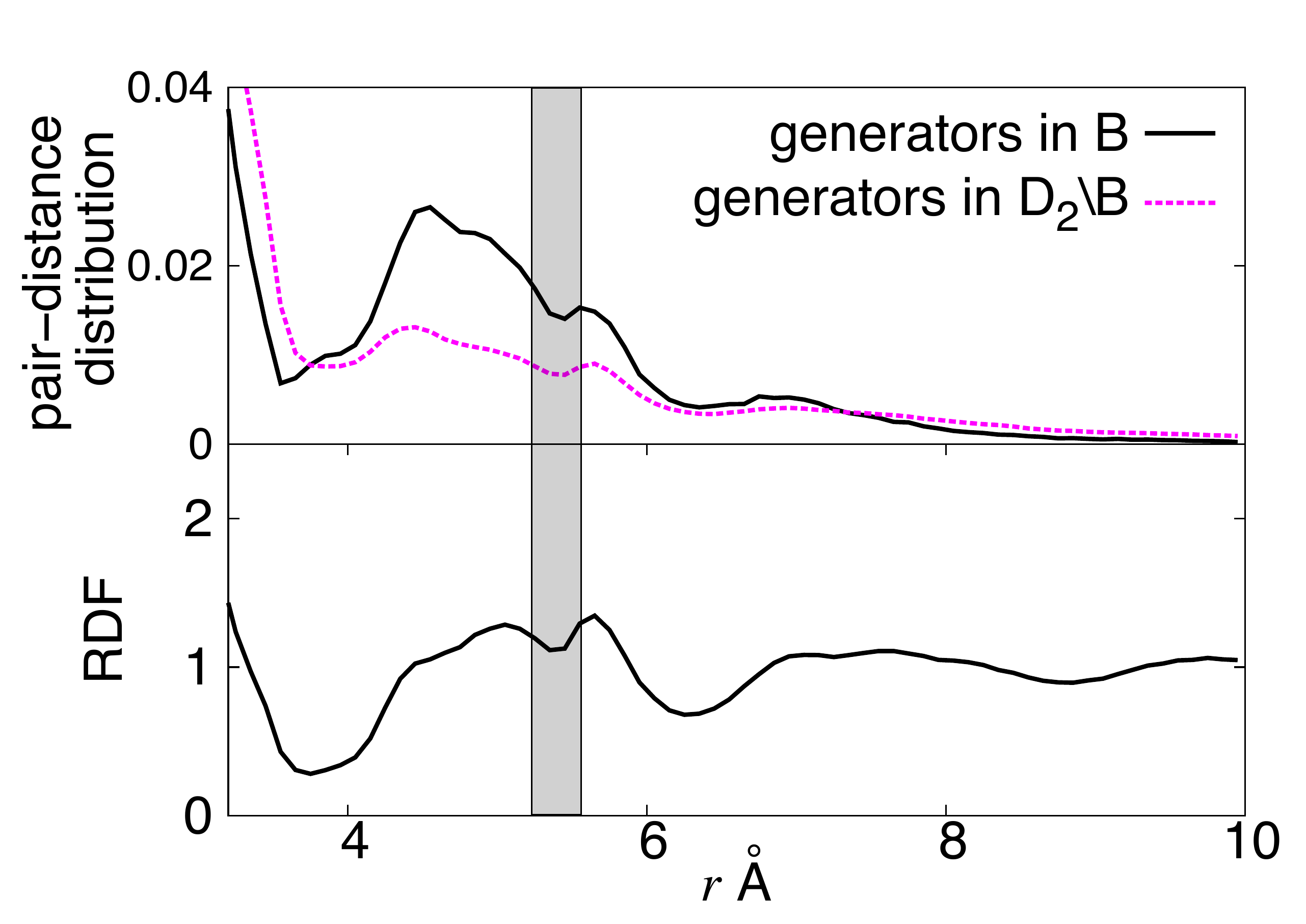}
\end{center}
\caption{The normalized pair-distance distributions (top panel) computed from $D_2(\cA^{\rm Cu_{50}Zr_{50}}_{\rm amo})$ and 
the radial distribution function for ${\rm Cu_{50}Zr_{50}}$ alloy (bottom panels) around the split second peak. 
The black line in the top panel was obtained by the pair-distances of the atoms 
in $B\subset D_2(\cA^{\rm Cu_{50}Zr_{50}}_{\rm amo})$,
whereas the pink line corresponds to those in the complement of $B$.
}
\label{optimal:gofr}
\end{figure}

In the random packing structure, the split second peak of the radial distribution function has been supposed to be a signature of MRO \cite{Barker}. 
The shaded region of the radial distribution function in Fig \ref{optimal:gofr} (bottom panel) shows the split second peak of ${\rm Cu_{50}Zr_{50}}$. 
Then, we found that the pair-distance distribution of the atoms of generators in 
$B:=[0.92,1.05]\times [1.76,2.01]\subset D_2(\cA^{\rm Cu_{50}Zr_{50}}_{\rm amo})$ also shows a clear splitting of the distribution (black line in the top panel of Fig. \ref{optimal:gofr}) in the same length-scale. 
Here, $B$ is chosen to be a region around the octahedral distribution. 
Meanwhile the pair-distance distribution of generators other than $B$ shows a slight change there (pink line in Fig. \ref{optimal:gofr}). 
This means that the generators around the octahedral distribution play a significant role for the split second peak. 
Therefore, similar to the FSDP in the silica, this result demonstrates that the PDs classify the length-scale of MRO from other scales.

We also studied the PDs of only the Zr component.
The PD $D_1$ for Zr in ${\rm Cu_{50}Zr_{50}}$  (middle left panel in Fig.~\ref{fig:CuZr_pd}) 
represents the existence of a hierarchical MRO structure similar to that of the silica in Fig.~\ref{fig:pd_cgl}, 
whereas the PD $D_1$ for Zr in ${\rm Cu_{15}Zr_{85}}$ (bottom left panel in Fig.~\ref{fig:CuZr_pd})
does not show any hierarchical curves. 
An example of the hierarchical rings in ${\rm Cu_{50}Zr_{50}}$ is depicted as the insets in the PD. 

We here remark that ${\rm Cu_{50}Zr_{50}}$ has higher glass forming ability than ${\rm Cu_{15}Zr_{85}}$ \cite{Ma2015}.   Relationships between the glass forming ability and the geometric structure are now actively studied (e.g., \cite{CM}).
Then, this result suggests another possibility that the presence of the hierarchical MRO structure extracted from PDs is also related to the glass forming ability of the alloy. 
In view of the results that the hierarchical MRO structure in PDs plays an important role for characterizing the glass states in ${\rm SiO_2}$, this statement seems to be reasonable. 
In order to understand the geometric and topological formation of the glass, it will be a great challenge to clarify this new perspective.

\section{Conclusion}
We have presented that  PD is a powerful tool for geometric characterizations of various amorphous solids in the short, medium, and even further ranges. 
In this work, we have addressed two different types of amorphous systems: continuous random network  and random packing structures. 
Both types of amorphous systems are characterized well by the existence of the curve and island distributions in the PDs. These specific distributions characterize the shapes of rings and cavities in multi-ranges, and the analysis using optimal cycles explicitly captures hierarchical structures of these shapes. 
We have shown that these shape characteristics successfully reproduce the FSDP for the continuous random network and the split second peak for the random packing, and provide further geometric insights to them. Furthermore, the global connectivity of dense packing regions in the Lennard-Jones system is revealed by the iterative application of the PD. 
For the binary random packing of the Cu-Zr metallic glass, we have also shown that the presence of the hierarchical MRO rings in the single component suggests the relationship with the glass forming ability.

The methodology presented here can be applied to a wide variety of disordered systems and enables one to survey the geometric features and constraints in seemingly random configurations. Furthermore, as we investigated the mechanical response of the PDs, studying dynamical properties of materials using the PD method would be of great importance to understand the relationship between hierarchical structures and mechanical properties. 
We believe that further developments and applications of topological data analysis will accelerate the understanding of amorphous solids.

\begin{acknowledgments}
Y.H. and T.N. contributed equally to this work. 
We thank Mingwei Chen, Hajime Tanaka, Masakazu Matsumoto, and Daniel Miles Packwood for valuable discussions and comments. 
This work was sponsored by ``WPI Research Center
Initiative for Atoms, Molecules and Materials,'' Ministry of
Education, Culture, Sports, Science and Technology of Japan.
Y.H. is supported by JST CREST Mathematics (15656429). 
Y.H, A.H., and Y.N. are supported by Structural Materials for Innovation D72, SIP.
K.M. is supported by ``MEXT, Coop With Math Program.'' 
Y.N. is supported by JSPS (26310205). 
T.N. is supported by JSPS (15K13530) and JST PRESTO. 
\end{acknowledgments}

{\bf \large Supporting Information: Hierarchical structures of amorphous solids characterized by persistent homology}

\section{Computational homology}
We summarize the computational homological tools used in this article. For further mathematical and computational details, please refer to [10].

\begin{figure}[h!]
\begin{center}
\includegraphics[height=3cm,width=3.5cm]{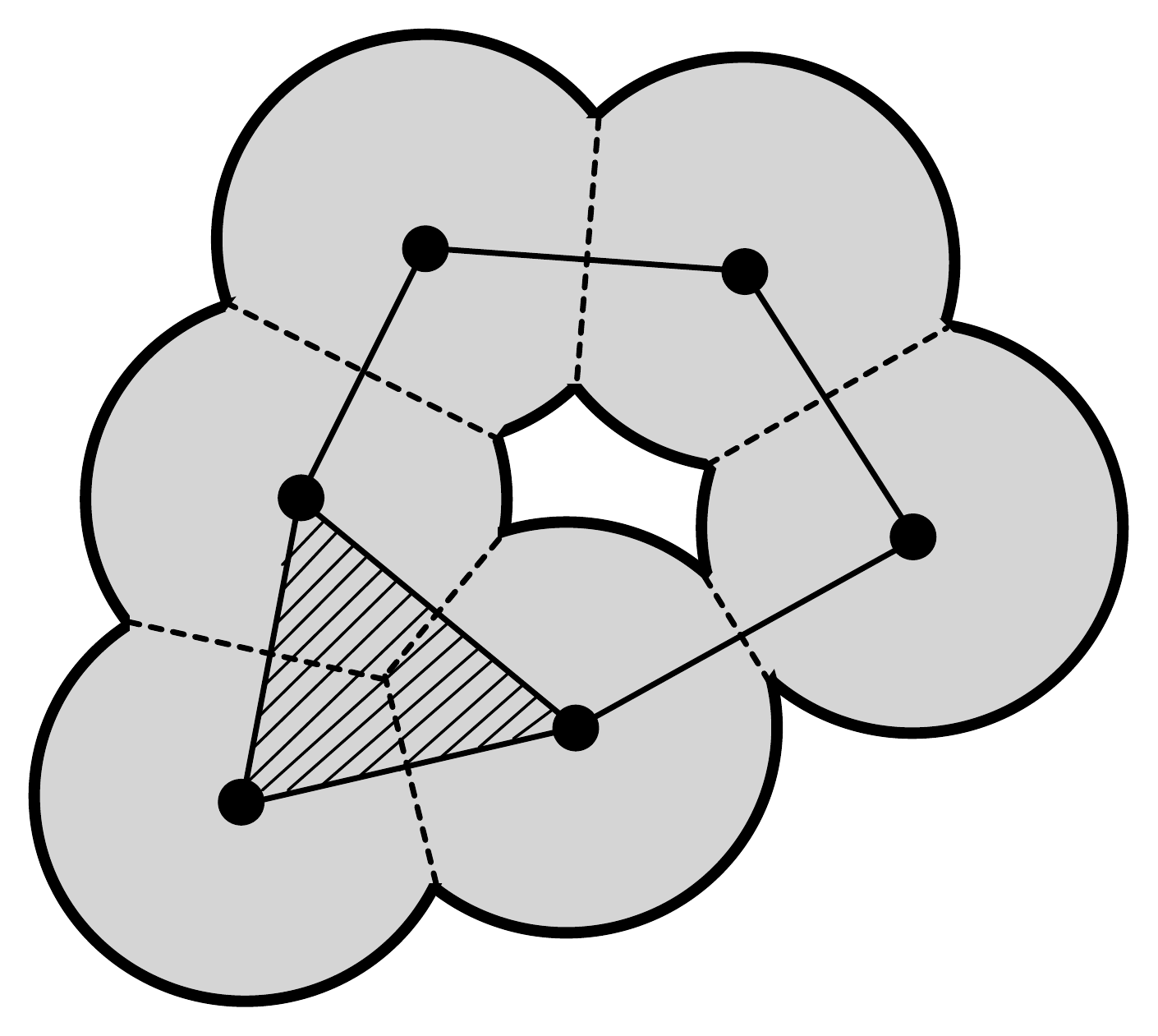}
\caption{Alpha shape.}
\label{fig:alpha}
\end{center}
\end{figure}

Given a pair $\mathcal{A}=(Q,R)$ of an atomic configuration $Q=(\bv x_1,\bv x_2\dots,\bv x_N)$ 
and a set $R=(r_1,r_2,\dots,r_N)$ of input radii, the persistence diagram captures hierarchical structures of the topological features such as rings and cavities in a one-parameter family of alpha shapes $A(Q,R;\alpha)$. Here, the alpha shape $A(Q,R;\alpha)$ is constructed as follows. Let $\mathbb{R}^3=\bigcup_{i=1}^NV_i$ be the Voronoi decomposition for $Q$, where $V_i$ is the Voronoi cell of $\bv x_i$. For each parameter $\alpha$, we assign a ball $B_i(\alpha)=\left\{ \bv x \in \R^3\mid ||\bv x-\bv x_i|| \le r_i(\alpha)\right\}$ with the radius $r_i(\alpha)=\sqrt{\alpha+r_i^2}$ for the $i$th atom. Then, the alpha shape $A(Q,R;\alpha)$ is defined by the dual of the cut balls $C_i(\alpha)=V_i\cap B_i(\alpha)$, $i=1,\dots,N$. Namely, this is a polyhedron with the vertex set $Q$ such that an edge $|{\bv x}_i {\bv x}_j|$ (triangle, tetrahedron, respectively) is assigned to $A(Q,R;\alpha)$ if and only if the intersection $C_i(\alpha)\bigcap C_j(\alpha)$ of the corresponding double (triple, quadruple, respectively) cut balls is nonempty (see Fig.~\ref{fig:alpha}).  In this article, the alpha shapes were computed using CGAL [20].

One of the important properties of the alpha shape is that the atomic ball model $B(\alpha)=\bigcup_{i=1}^NB_i(\alpha)$ and the alpha shape $A(Q,R;\alpha)$ can be continuously deformed to each other. Hence, there is no loss of topological information in the use of the alpha shape model, which is much easier to analyze computationally than the atomic ball model. Furthermore, we can trace the hierarchical structures in $B(\alpha)$ by changing the parameter $\alpha\geq\alpha_{\rm min}$, where $\alpha_{\rm min}=-\min\{r_1^2,\dots r_N^2\}$. In other words, the parameter $\alpha$ controls the resolution from fine ($\alpha\approx \alpha_{\rm min}$) to crude ($\alpha\gg \alpha_{\rm min}$) geometric features.

\begin{figure*}[t!]
\begin{center}
\includegraphics[height=4.8cm,width=16cm]{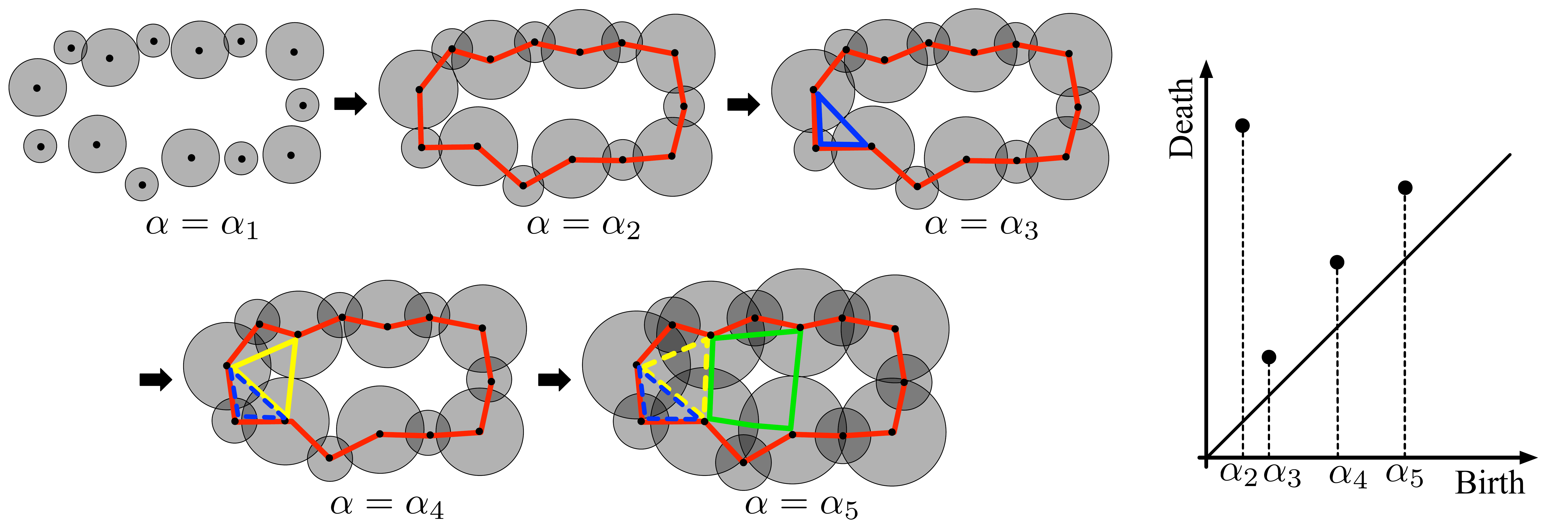}
\caption{Schematic illustration of atomic ball models $B(\alpha)$ showing the rings on $C_{\rm P}$ (red), $C_{\rm T}$ (blue), $C_{\rm O}$ (yellow), and $B_{\rm O}$ (green). Here, the large (small) balls correspond to oxygen (silicon). As we increase the parameter $\alpha$, the primary ring  generates the secondary rings in the hierarchy.}
\label{fig:point_cloud}
\end{center}
\end{figure*}

The left of Fig.~\ref{fig:point_cloud} shows a schematic illustration of atomic ball models $B(\alpha)$ with the rings on $C_{\rm P}$ (red), $C_{\rm T}$ (blue), $C_{\rm O}$ (yellow), and $B_{\rm O}$ (green). 
Here, the large (small) balls correspond to oxygen (silicon).
At each $\alpha_i$ ($i=2,3,4,5$), a new ring appears, and the dashed rings express the rings that have disappeared. The hierarchical ring structures are observed at $\alpha_3$, $\alpha_4$, and $\alpha_5$, at which new rings are generated by pinching the primary ring.

The persistence diagram is a two-dimensional histogram recording the birth and death scales of the topological features in a one-parameter family $\{B(\alpha)\}_{\alpha}$ of atomic ball models. For example, let $c$ be the yellow ring in Fig.~\ref{fig:point_cloud}. We observe that $c$ first appears at $\alpha=\alpha_4$ and first disappears at some value $\alpha_*$, where $\alpha_4<\alpha_*<\alpha_5$. Then, the birth and death scales $b$ and $d$, respectively, of the ring $c$ are determined by $b=\alpha_4$ and $d=\alpha_*$.  
The histogram of the birth and death pairs $(b,d)\in\mathbb{R}^2$ for all the rings appearing in the one-parameter family $\{B(\alpha)\}_{\alpha}$ is called the persistence diagram $D_1(\mathcal{A})$. The right of Fig.~\ref{fig:point_cloud} shows the persistence diagram of the left. 
From the definition of the persistence diagram, the birth scale indicates the maximum distance between adjacent atoms in the ring, whereas the death scale indicates the size of the ring. 
It should also be mentioned that persistence diagrams $D_i(\mathcal{A})$ can be similarly defined for any nonnegative integer $i\in\{0,1,2,\dots\}$, and they capture higher-dimensional topological features e.g., cavities for $i=2$. The persistence diagrams in this article were computed by PHAT [21].

\section{MD simulation for the silica system} 
The atomic configurations of silica in a liquid $Q_{\rm liq}$, 
an amorphous solid $Q_{\rm amo}$, and a crystalline solid $Q_{\rm cry}$ state were generated by the MD simulation.
The system was composed of 2,700 silicon atoms and 5,400 oxygen atoms
interacting via the BKS potential [22],
and the 24-6 Lennard-Jones potential is assigned 
to correct for the unphysical behavior during equilibration in the liquid state [30].
The masses of the silicon and oxygen atoms were 27.98 and 15.99 grams/mole respectively.
Despite its simplicity, the BKS model reasonably reproduces the static and dynamic properties of amorphous silica 
[31].

Starting from the beta-cristobalite structure, 
the crystalline configuration $Q_{\rm cry}$ was obtained after the equilibration 
for 5 ps at 1 bar and 10 K.
The MD simulation was performed with 
Parrinello-Raman barostat and Nos\'e-Hoover thermostat 
with the damping parameters 
$\tau_{\rm bar}=0.5$ ps and $\tau_{\rm th}=0.8$ ps, respectively.
The time step was set to be 1.0 fs.

The liquid configuration $Q_{\rm liq}$ was obtained after the equilibrating the system 
from a random initial configuration for 50 ps at 1 bar and 7,000 K.
Here, the MD simulation was performed with the Andersen barostat and 
the Nos\'e-Hoover thermostat.

The amorphous configuration $Q_{\rm amo}$ was obtained after 
the cooling MD simulation [31].
Starting from the liquid configuration at the temperature 7,000 K, 
the heat-bath temperature was decreased linearly 
to the final temperature 10 K with $-dT/dt=10^{12}$ K/s.
Throughout the cooling process, the pressure was kept at 1 bar
and we used the Nos\'e-Hoover thermostat with $\tau_{\rm th}=1.0$ ps.
In the high temperatures region 1,500 K $\le T \le$  6,000 K,
we used Anderson barostat with $\tau_{\rm bar}=1.0$ ps.
In the low temperatures region 10 K $\le T \le$ 1,500 K,
we used the Parrinello-Raman barostat with $\tau_{\rm bar}=0.5$ ps.
Then, $Q_{\rm amo}$ was obtained as the final configuration of the cooling process.
We tested five cooling rates, $-dT/dt=10^{15}, 10^{14}, 10^{13}, 10^{12}$, $10^{11}$ K/s, and checked that these choices do not affect our results qualitatively.

The set $R=(r_1,\cdots,r_N)$ of input radii can be adjusted for each $i$th atom ($i=1,\dots,N$). 
In our analysis, we chose the radii $r_{\rm Si}$ and $r_{\rm O}$ 
of silicon and oxygen using the first peak positions of 
the partial radial distribution functions of $Q_{\rm amo}$. 
The first peaks for SiO, OO, and SiSi appeared at $r_{\rm SiO}=1.65$ \AA, 
$r_{\rm OO}=2.55$ \AA, and $r_{\rm SiSi}=3.15$ \AA, respectively. 
Then, we selected the smallest two peaks and determined $r_{\rm Si}$ and 
$r_{\rm O}$ by solving $r_{\rm Si}+r_{\rm O}=r_{\rm SiO}$ and $2 r_{\rm O}=r_{\rm OO}$. 
This led to $r_{\rm O}=1.275$ \AA~and $r_{\rm Si}=0.375$ \AA. 

\begin{figure}[h!]
\begin{center}
\includegraphics[height=5.0cm,width=8cm]{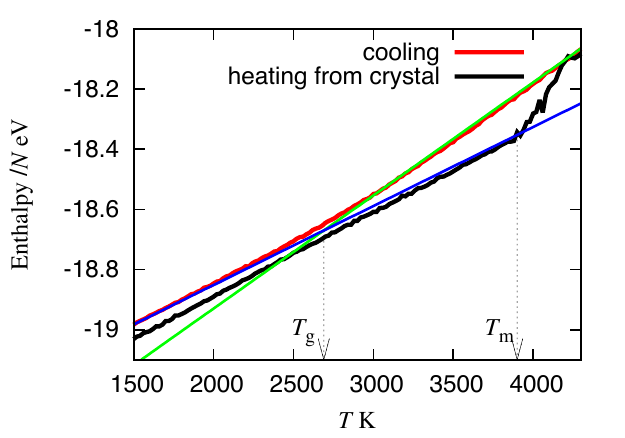}
\caption{Enthalpy as a function of temperature for cooling 
and heating of silica system.}
\label{fig:TgTm}
\end{center}
\end{figure}

\begin{figure}[h!]
\includegraphics[height=3.3cm,width=8.3cm]{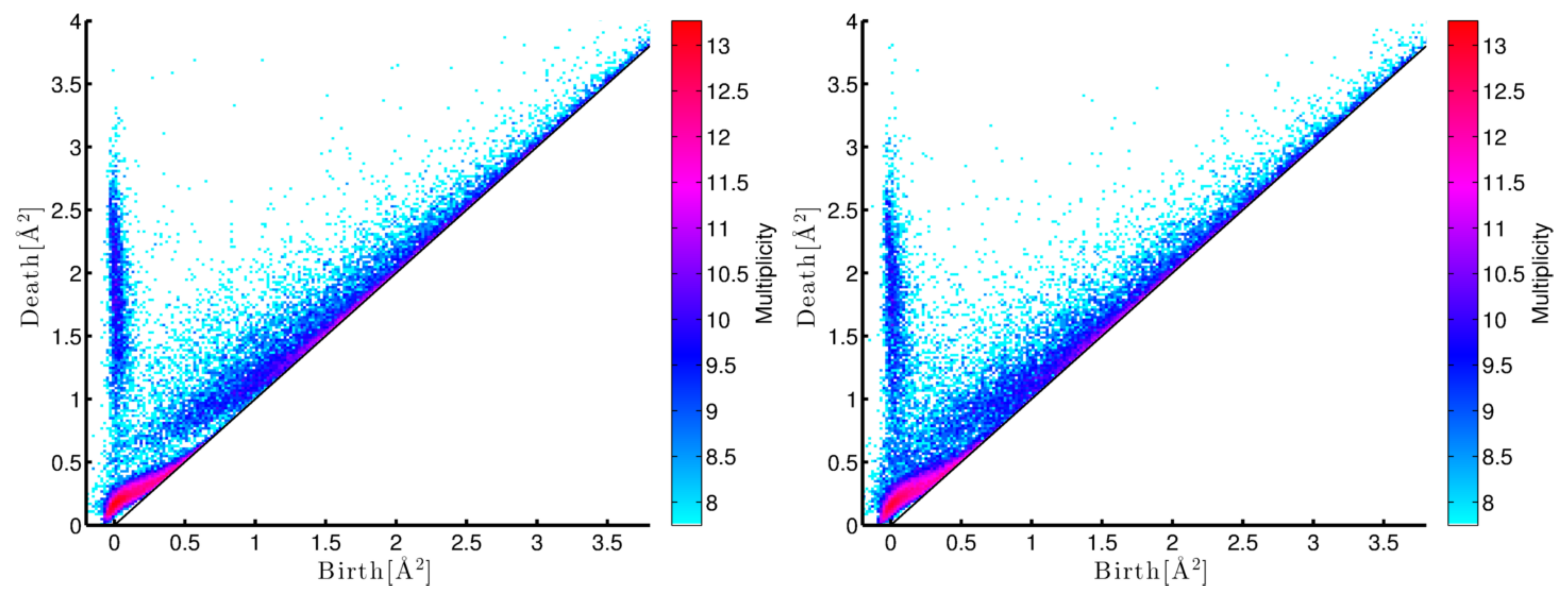}
\caption{$D_1(\cA)$
below $T_{\rm g}$ (left) and in the super-cooled liquid region (right).}
\label{fig:PDTgTm}
\end{figure}

The glass transition temperature $T_{\rm g}$ and 
the melting temperature $T_{\rm m}$ are
estimated through the enthalpy temperature curves.
Starting from beta-cristobalite configuration,
the system is heated with a heating rate $dT/dt=0.1$ K/ps
from $T=1$ K to $T=7,000$K (black curve in Fig.~\ref{fig:TgTm}).
Then, $T_{\rm m}$ is evaluated to be around $3,900$K [32].
Subsequently, the cooling simulation is performed 
with a cooling rate $-dT/dt=10.0$ K/ps (red curve).
Then the glass transition temperature is evaluated  
as an intersection point between tangential lines of 
low (blue line) and high (green line) temperature regions of the curve
and found to be $T_{\rm g}\approx 2,700$ K [31].
Fig.~\ref{fig:PDTgTm} shows PDs for $2,500$ K (left) and $T=3,500$ K (right), 
and the curve $C_{\rm O}$
starts to appear below $T_{\rm g}$.

\section{MD simulation for the LJ system and the CuZr alloy}
The Lennard-Jones system
is composed of 4,000 monatomic particles 
interacting via Lennard-Jones (LJ) potential 
$u(r)=4\varepsilon( (\sigma/r)^{12}-(\sigma/r)^{6})$
with $\varepsilon$, $\sigma$ and mass equal to unity.
The time step is set to be 0.005 in LJ units.

\begin{figure}[h!]
\includegraphics[height=4cm,width=8.8cm]{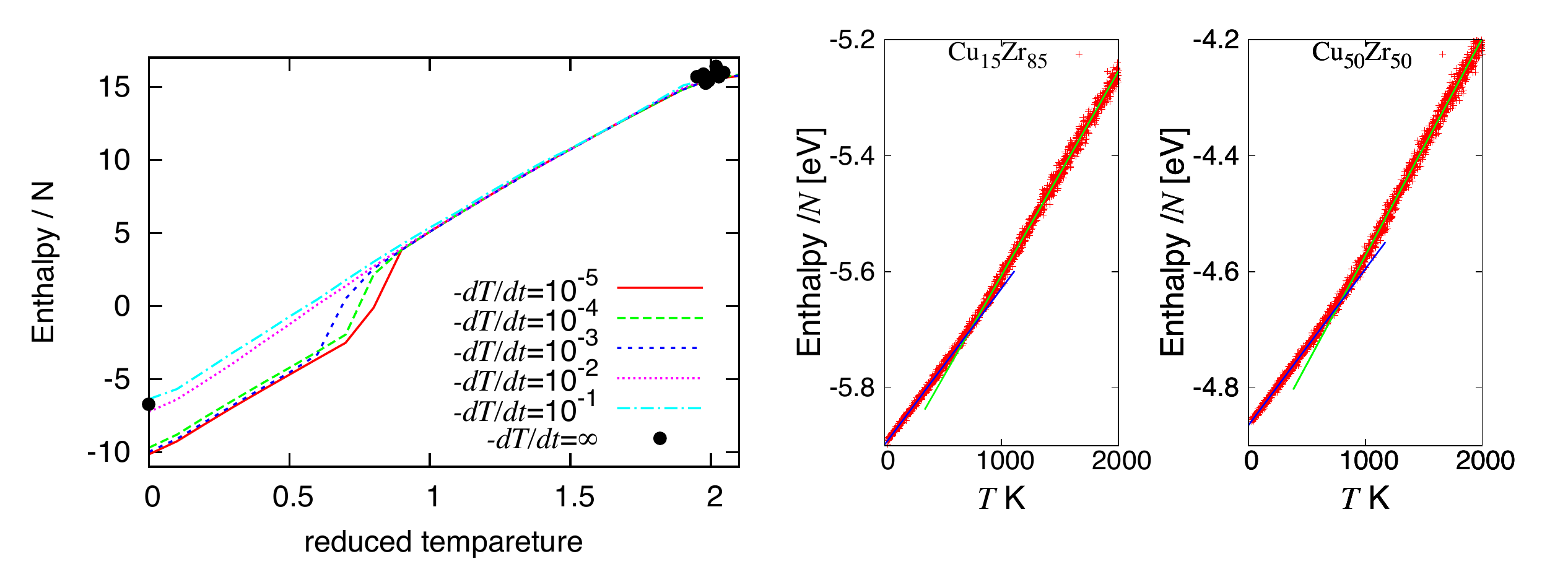}
\caption{Enthalpy as a function of temperature during cooling processes
for Lennard-Jones system (left) and CuZr alloy (right), respectively.}
\label{fig:EnthalpyTg}
\end{figure}

Starting from the FCC configuration, 
$Q^{\rm LJ}_{\rm cry}$ is obtained after equilibration 
at temperature $T=0.1$ and pressure $p=1$ 
with $\tau_{\rm th}=1.0$ and $\tau_{\rm bar}=1.0$.
To obtain an amorphous structure $Q^{\rm LJ}_{\rm amo}$, 
the equilibration at $T=2$ in the normal liquid phase
has been performed for 2000 steps 
at the constant number density 1.015 [33]
with $\tau_{\rm th}=0.5$.
Then, the cooling MD simulations have been performed 
for several cooling ratios from $-dT/dt=10^{-5}$ to $10^{-1}$.
In the left panel in Fig.~\ref{fig:EnthalpyTg},
the enthalpy normalized by the number of particles is described as a function of $T$.
As is observed, there are no glass transition in the LJ system.
Therefore, we use an inherent structure for the liquid state as $Q^{\rm LJ}_{\rm amo}$
(black points in the left panel in Fig.~\ref{fig:EnthalpyTg}).

For the Cu-Zr alloy system,
MD simulations have been performed 
using the embedded-atom method potential [34].
The masses of Cu and Zr atoms were set to be 63.54 and 91.22 grams/mole, respectively.
The number of particles was 16,000, 
in which there were 8,000 of both copper and zirconium atoms for ${\rm Cu_{50}Zr_{50}}$, 
and 2,400 were copper atoms and 13,600 were zirconium atoms for ${\rm Cu_{15}Zr_{85}}$.

In this case, we can observe the glass transition 
(the right panel in Fig.~\ref{fig:EnthalpyTg})
during a cooling simulation with the quench rate $10^{13}$ K/sec [26] 
down to $T=1$ K keeping the pressure $p=0$ atm after the equilibration at $T=17,000$ K.
Then, we use the final configuration at $T=1$ K as the amorphous configuration.

All MD simulations in this article were performed using LAMMPS [35].
\end{article}

\end{document}